\documentclass[lettersize,journal]{IEEEtran}
\usepackage{amsmath,amsfonts}
\usepackage{algorithmic}
\usepackage{algorithm}
\usepackage{array}
\usepackage[caption=false,font=normalsize,labelfont=sf,textfont=sf]{subfig}
\usepackage{textcomp}
\usepackage{stfloats}
\usepackage{url}
\usepackage{verbatim}
\usepackage{graphicx}
\usepackage{cite}
\usepackage{xcolor}
\begin{document}

\title{Towards novel tunability schemes for hybrid ferromagnetic transmon qubits}

\author{Halima Giovanna Ahmad,~\IEEEmembership{Member,~IEEE,}, Raffaella Ferraiuolo*, Giuseppe Serpico*, Roberta Satariano, Anna Levochkina, Antonio Vettoliere, Carmine Granata, Domenico Montemurro, Martina Esposito, Giovanni Ausanio, Loredana Parlato, Giovanni Piero Pepe~\IEEEmembership{Member,~IEEE,}, Alessandro Bruno, Francesco Tafuri, Davide Massarotti
\thanks{The work was supported by the Pathfinder EIC 2023 project "FERROMON-Ferrotransmons and Ferrogatemons for Scalable Superconducting Quantum Computers"; the PNRR MUR project PE0000023-NQSTI and the PNRR MUR project CN 00000013-ICSC.
H.G.A., D.M. (Davide Massarotti) and F.T. thank SUPERQUMAP
project (COST Action CA21144).}
\thanks{* These authors have equally contributed to the manuscript.}
\thanks{H.G.Ahmad, G.Serpico and F. Tafuri are with Dipartimento di Fisica "Ettore Pancini", Università degli Studi di Napoli "Federico II", Via Cinthia, 80126, Napoli, IT.}
\thanks{R.Ferraiuolo and A.Bruno are with QuantWare, Elektronicaweg 10, 2628 XG Delft, The Netherlands}
\thanks{R.Satariano, A. Levochkina, D.Montemurro, G.Ausanio, L.Parlato, G.P.Pepe are with Dipartimento di Fisica "Ettore Pancini", Università degli Studi di Napoli "Federico II", Via Cinthia, 80126, Napoli, IT and Consiglio Nazionale delle Ricerche - SPIN, c/o Complesso Monte Sant’Angelo, via Cinthia 26, I-80126, Napoli, IT. }
\thanks{M. Esposito is with Consiglio Nazionale delle Ricerche - SPIN, c/o Complesso Monte Sant’Angelo, via Cinthia 26, I-80126, Napoli, IT}
\thanks{A.Vettoliere and C.Granata are with Consiglio Nazionale delle Ricerche-ISASI, Via Campi Flegrei 34, I-80078, Pozzuoli, IT
}
\thanks{D.Massarotti is with Dipartimento di Ingegneria Elettrica e delle Tecnologie dell’Informazione, Universitá degli Studi di Napoli Federico II, Via Claudio 21, 80125, Napoli, IT and Consiglio Nazionale delle Ricerche - SPIN, c/o Complesso Monte Sant’Angelo, via Cinthia 26, I-80126, Napoli, IT}
}



\maketitle

\begin{abstract}

Flux tuning of qubit frequencies in superconducting quantum processors is fundamental for implementing single and multi-qubit gates in quantum algorithms. Typical architectures involve the use of DC or fast RF lines. However, these lines introduce significant heat dissipation and undesirable decoherence mechanisms,  leading to a severe bottleneck for scalability.  Among different solutions to overcome this issue, we propose integrating tunnel Superconductor-Insulating-thin superconducting interlayer-Ferromagnet-Superconductor Josephson junctions (SIsFS JJs) into a novel transmon qubit design, the so-called ferrotransmon. SIsFS JJs provide memory properties due to the presence of ferromagnetic barriers and preserve at the same time the low-dissipative behavior of tunnel-insulating JJs, thus promoting an alternative tuning of the qubit frequency. In this work, we discuss the fundamental steps towards the implementation of this hybrid ferromagnetic transmon. We will give a special focus  on the design, simulations, and preliminary experimental characterization of superconducting lines to provide in-plane magnetic fields, fundamental for an on-chip control of the qubit frequencies in the ferrotransmon.

\end{abstract}

\begin{IEEEkeywords}
Ferromagnetic Josephson junctions, quantum circuits, superconducting qubits, electromagnetic simulations
\end{IEEEkeywords}

\section{Introduction}

The Josephson effect is the backbone of superconducting quantum technologies. The non-linear potential energy of Josephson junctions (JJs) allows for engineering on-chip macroscopic artificial atoms ~\cite{tafuri2019} with transition energies that can be defined by the materials and circuit design~\cite{Koch2007,Clarke2008,DiCarlo2009,Devoret2013,Kockum2019,Krantz2019,Bal2024}. The capability to couple JJs with several circuital elements, such as transmission lines and superconducting resonators~\cite{Koch2007,DiCarlo2009,Manucharyan2009,Chen2014,Krantz2019,Kjaergaard2020,Feofanov2010,Oliver2013,Lee2019, Place2021}, provides an easy interface with control and readout electronics, which is fundamental for basic quantum computing operations and for fulfilling the DiVincenzo's criteria~\cite{DiVincenzo}.  Most  importantly, the noticeable advances in material science,  nanofabrication techniques and circuit design have led to a large variety of superconducting qubits, each of them with different performances in terms of coherence and fidelities~\cite{Siddiqi2021}. Therefore, the search for combinations of novel materials, circuital designs, and new protocols for encoding qubits has emerged as a very active field\cite{DiCarlo2009,Feofanov2010,Oliver2013,Chen2014,McDermott2018,Mukhanov2019,Krantz2019,Lee2019,Kjaergaard2020,Place2021,Manucharyan2009,Bal2024}. 

It is a common request in transmon devices to have an in-situ knob to tune the qubit frequency. This is usually achieved by integrating Josephson DC-SQUIDs~\cite{Barone1982,Tafuri2003,Koch2007,Hutchings2017}, where the Josephson energy can be tuned by threading an external magnetic field through the DC-SQUID loop. Such magnetic flux fields are typically generated on-chip by inductively coupling DC or fast RF transmission lines~\cite{Koch2007,Hutchings2017}. By sending DC or AC signals through flux lines, it is possible to: i) set the idle frequency of the qubits in superconducting quantum processing units for certain specific operations~\cite{Asaad2016,krinner2022}; ii) implement single-qubit Z quantum gates~\cite{Krantz2019} and iii) set on resonance two or more coupled qubits to implement two-qubit entangling gates, like the iSWAP or the CZ~\cite{DiCarlo2009,Yuan2020,Sung2021,Ahmad2024Q}. Nevertheless, flux tunability poses important limitations in scalability and coherence~\cite{Koch2007,kempf2016,Hutchings2017,Quintana2017,Ahmad2023}. Indeed, magnetic flux fluctuations are one of the primary noise sources in transmon qubits~\cite{Koch2007,Yan2016}, and solutions that reduce their contribution are required. To reach this goal, unconventional hybrid JJs have been proposed to achieve novel tunability schemes in superconducting qubits. Superconducting-semiconducting hybrid JJs, for example, have been integrated into transmon qubits to achieve qubit frequency tunability using gate voltages (gatemon), rather than magnetic flux fields\cite{Barthel2009,Larsen2015,DeLange2015,Wiedenmann2016,Manousakis2017,Gul2018,Kroll2018,Casparis2018,Kunakova2020}. 
\begin{figure}[t]
    \centering
    \includegraphics[width=0.8\linewidth]{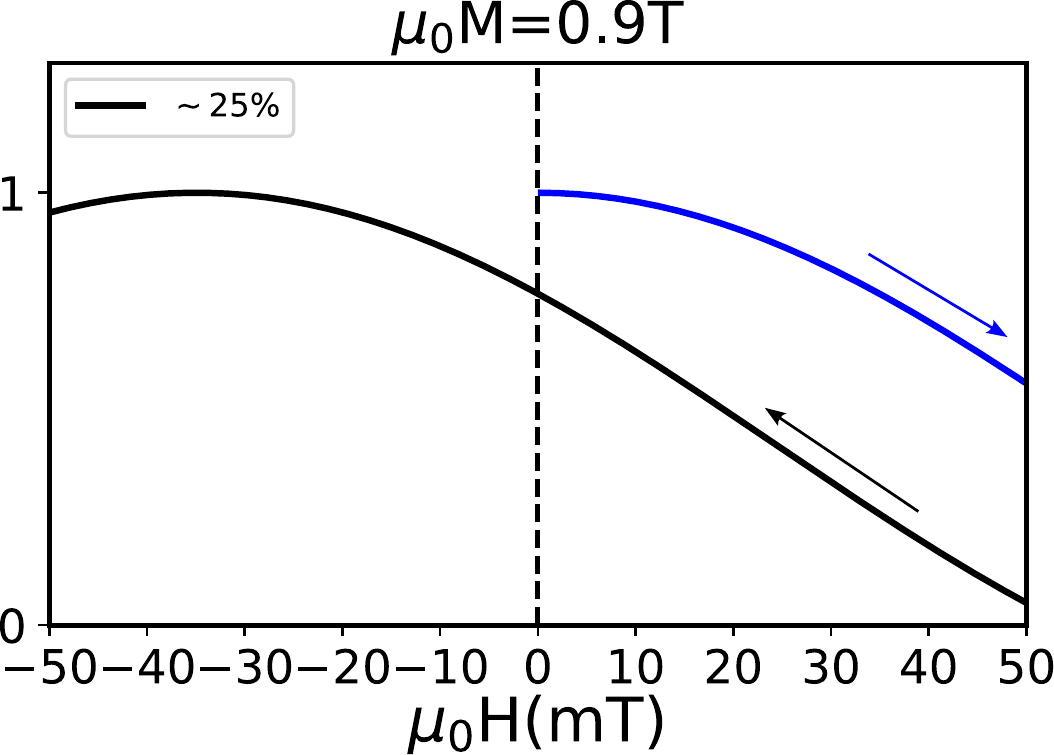}
    \caption{Simulation for normalized Fraunhofer patterns of a square SIsFS Josephson junction with side length $a = 300\;nm$. The blue and black curves identify the Fraunhofer pattern corresponding to the virgin magnetization curve and the Fraunhofer pattern obtained by ramping the external magnetic field in the downward direction, as shown by the arrow. The critical current level separation at the zero-field working point is $\sim 25\%$. The simulation considers a saturation magnetization $\mu_0M=0.9\;T$.}
    \label{fig:pattern_quad_rettang}
\end{figure}

In Ref.~\cite{Ahmad2022}, we proposed to integrate hybrid tunnel ferromagnetic JJs in transmon devices (ferrotransmon) to provide a novel frequency tuning scheme that relies on the hysteretic magnetic behavior of the ferromagnetic barrier~\cite{Larkin2012,Blamire2013,Vernik2013}: by applying an in-plane magnetic field, perpendicular to the Josephson current, a hysteretic Fraunhofer-like behavior of the critical current $I_c$ is expected with a shift depending on the sweeping magnetic field direction. In Figure~\ref{fig:pattern_quad_rettang}, we report a simulation of the normalized Fraunhofer-like modulation for a tunnel ferromagnetic JJ with a lateral size in the submicrometer regime, compatible with transmon circuits~\cite{Krantz2019}. Starting from the virgin magnetization curve, i.\,e., when the ferromagnetic barrier has not yet been magnetized, the magnetic field is swept towards positive values on the rise of the in-plane magnetic field pulse (in blue). In contrast, on the falling edge of the pulse (in black), the ferromagnetic barrier acquires a positive residual magnetization, thus effectively shifting the $I_c$ Fraunhofer-like modulation to negative magnetic fields. Therefore, the JJ will show a different $I_c$ value at zero-field, i.\,e., effectively switching off the external magnetic fields while performing quantum logic operations. A similar protocol has been earlier proposed for the implementation of cryogenic memories~\cite{Weides2006,Larkin2012,Vernik2013,Caruso2018,Parlato2020,Ryazanov2021}, where a fundamental requirement was to achieve sufficiently high switching speed between the current level states. In this sense, tunnel ferromagnetic JJs have been demonstrated to achieve faster switching speed compared to standard metallic ferromagnetic JJs~\cite{Weides2006,Larkin2012,Vernik2013,Caruso2018,Parlato2020,Ryazanov2021}: the fundamental timescale is given by the Josephson switching speed $\tau\propto I_cR_N$, where the normal resistance $R_N$ of a tunnel JJ is orders of magnitudes larger than in metallic JJs~\cite{Weides2006,Larkin2012,Vernik2013,Caruso2018,Parlato2020,Ryazanov2021}. Specifically for what concerns quantum technologies applications, tunnel-like behavior of the JJ is fundamental also to achieve an underdamped electrodynamics behavior, i.\,e., to guarantee a sufficiently low impact of quasiparticle dissipation~\cite{Massarotti2015,Ahmad2020,Ahmad2024}. 

In this manuscript, we first summarize the main requirements to build a proof-of-concept hardware for the ferrotransmon, discussed in detail in previous works~\cite{Ahmad2022,vettoliere2022,Vettoliere2022aluminum,Ahmad2024,Satariano2024,Satariano2024low}. In Sec.~\ref{Design}, we will then focus on our proposals for on-chip fast in-plane magnetic field lines  layout, a fundamental step towards the experimental validation of the ferrotransmon design. We will finally outline the conclusions and next steps to follow shortly.

\section{Experimental requirements towards the ferrotransmon}

The first fundamental requirement towards the integration of tunnel ferromagnetic JJs in transmon devices is that these systems must be compatible with common fabrication techniques and materials employed for the qubits~\cite{vettoliere2022,Vettoliere2022aluminum,Ahmad2024,Satariano2024}. Most of the transmon technology uses aluminum SIS JJs, with critical current of tens of nanoamperes~\cite{Krantz2019}, standard BCS-like transport, and noticeable performances in terms of quasiparticles dissipation and quality factors~\cite{Devoret1985,Martinis1987}. The second request is flexibility regarding the choice of ferromagnetic materials to integrate into the JJ~\cite{Ahmad2020}. For this reason, we have focused our attention in the last years on Superconductor-Insulating-thin superconducting interlayer-Ferromagnet-Superconductor (SIsFS) JJs~\cite{Parlato2020,Satariano2021,vettoliere2022,Vettoliere2022aluminum,Ahmad2024,Satariano2024}.

SIsFS JJs have the advantage of showing different transport regimes as a function of materials and  thickness of  the layers in the structure~\cite{Bakurskiy2013,arxivBakurski}. 
To fall into the tunnel limit, the thickness of the superconducting interlayer \(d_{\mathrm{s}}\) has to be larger than  the critical thickness \textit{d}\(_{\mathrm{sc}}\), i.e., the minimal thickness of the s layer in a sF bilayer above which superconductivity still exists at a certain temperature. With this condition, the pair potential \(\Delta\), i.\,e. the energy gap of the superconducting s layer, is close to that of the bulk material and the SIsFS structure behaves as a serial connection of a tunnel SIs JJ and a ferromagnetic sFS JJ ~\cite{Bakurskiy2013,arxivBakurski,vettoliere2022,Satariano2024}. For small F-thickness \textit{d}\(_{\mathrm{F}}\), because of the metallic nature of standard F barrier and resulting higher barrier transparency, we get ${I_{c}}^{sFS}\gg{I_{c}}^{SIs}$~\cite{Buzdin1991}, thus ensuring that dissipative effects are regulated entirely by the SIs counterpart, without any detrimental effect from the diffusive and metallic sFS JJ. 
This has been demonstrated through different transport experiments at dilution temperatures on SIsFS JJs based on aluminum technology in Refs.~\cite{vettoliere2022,Vettoliere2022aluminum,Ahmad2024}. A comparison with similar SIsS JJs has shown no suppression of the superconducting gap, expected in the case of a non-series transport~\cite{vettoliere2022,Vettoliere2022aluminum}, and no detrimental impact on the subgap branch due to the presence of the ferromagnet~\cite{Ahmad2024}. The electrodynamic parameters of the analyzed JJs are also compatible with typical transmon designs~\cite{Ahmad2020,vettoliere2022}, and show no deviations compared to reference samples without the ferromagnetic interlayer.

At the same time, one can regulate the behavior in the presence of in-plane magnetic fields of a SIsFS JJ by making $d_s$ smaller than the London penetration depth $\lambda_s$~\cite{Volkov2019,Dahir2019,Satariano2024}. If the condition $d_s<\lambda_s$ is satisfied, the SIsFS JJ behaves as a single junction in the presence of an in-plane magnetic field. 

The Fraunhofer-like $I_c(H)$ modulation of an SIsFS JJ is influenced by two key parameters: the area of the JJ and the ferromagnetic barrier. The former sets the width of the first lobe of the pattern and depends on the magnetic area and geometry of the device: the larger it is, the smaller the required on-chip magnetic fields to tune the $I_c$~\cite{Barone1982}. However, typical areas for transmon circuits fall in the submicrometer scale to achieve suitable Josephson and charging energy~\cite{Koch2007}. In  Figure~\ref{fig:pattern_quad_rettang}, for example, we have simulated the Fraunhofer-like $I_c(H)$ curve for a square SIsFS JJ with lateral size of $300\;nm$. Here, the Fraunhofer pattern first lobe width allows to achieve a critical current level separation between $20-30\%$ by applying magnetic field pulses with amplitude of the order of $5\;mT$. 

The choice of the ferromagnetic layer becomes then crucial since it affects the hysteresis of the $I_c(H)$ curve~\cite{Robinson2012,Blamire2013,Blamire2014,Satariano2021,Satariano2024}: for JJs with the same geometry, the largest the residual magnetization of the ferromagnet, the greater is the critical current level separation. So far, SIsFS JJs have used soft weak ferromagnets, such as PdFe barriers, which offer on one hand very small coercive fields~\cite{Larkin2012,Caruso2018,Karelina2021}. On the other hand, PdFe is unsuitable for devices with reduced areas due to the percolative nature of the exchange interaction between iron atoms~\cite{Bolginov2017}, which can lead to frustrated magnetic properties when the dimensions are scaled down. To scale JJs down to submicron dimensions, strong ferromagnets with high remanent magnetization and in-plane anisotropy are required to distinguish between the two critical current states. In our previous experiments, we used superconducting coils to demonstrate the magnetic behavior of SIsFS JJs based on Fe$_{20}$ Ni$_{80}$ alloy (permalloy: Py) through Josephson magnetometry~\cite{Parlato2020,Satariano2021,vettoliere2022,Vettoliere2022aluminum,Ahmad2024}. These experiments showed that Py exhibits desirable switching properties with relatively low coercive fields even at smaller sizes~\cite{Niedzielski_2015, Qader2014_small_magnetic_fields_switching}. As demonstrated in Ref.~\cite{Parlato2020}, Py performs well when scaled down to the micron regime, with important implications for scalability. 
\begin{figure}
\centering	\includegraphics[width=0.9\columnwidth]{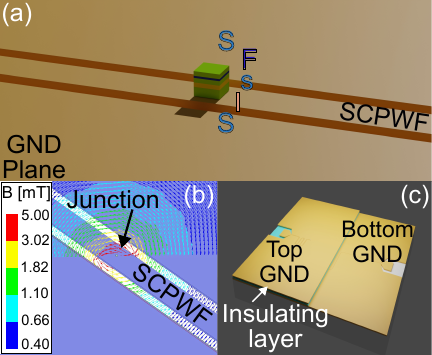}

	\caption{(a) 3D schematic (not to scale) of the SIsFS Josephson junction above the superconducting coplanar waveguide 
	flux line. (b) Ansys simulation results show the magnetic field distribution generated by the CPW flux line, highlighting the uniformity of the magnetic field in the junction area. (c) Sample cross-sectional configuration featuring a bottom ground plane and a top ground plane separated by an insulating layer, allowing for effective bonding of both grounds.}
	\label{fig_SCPWF}     
\end{figure}

Although the use of coils did not pose limitations in the experimental investigation of the former devices, these are not well-suited for scalable quantum processing units. Applying a magnetic field with a coil would simultaneously affect all qubits, which is undesirable. Each qubit must have a dedicated line for localized tunability. Therefore, a fundamental step towards the ferrotransmon validation and proof-of-principle is to search for circuital solutions able to generate magnetic fields in-plane compatible with both dimensions and magnetic strength parameters. This is not a common request, since the direction of the magnetic field for flux-tunable transmons is typically perpendicular to the chip,  and achieved by designing a superconducting $50\;\Omega$-matched transmission line inductively coupled to the transmon DC-SQUID. Hence, we here discuss possible solutions for the integration of in-plane magnetic fields generated through on-chip lines.

\section{On-chip fast in-plane magnetic field lines proposals}
\label{Design}

\begin{figure}[t]
	\centering
	\includegraphics[width=0.7\columnwidth]{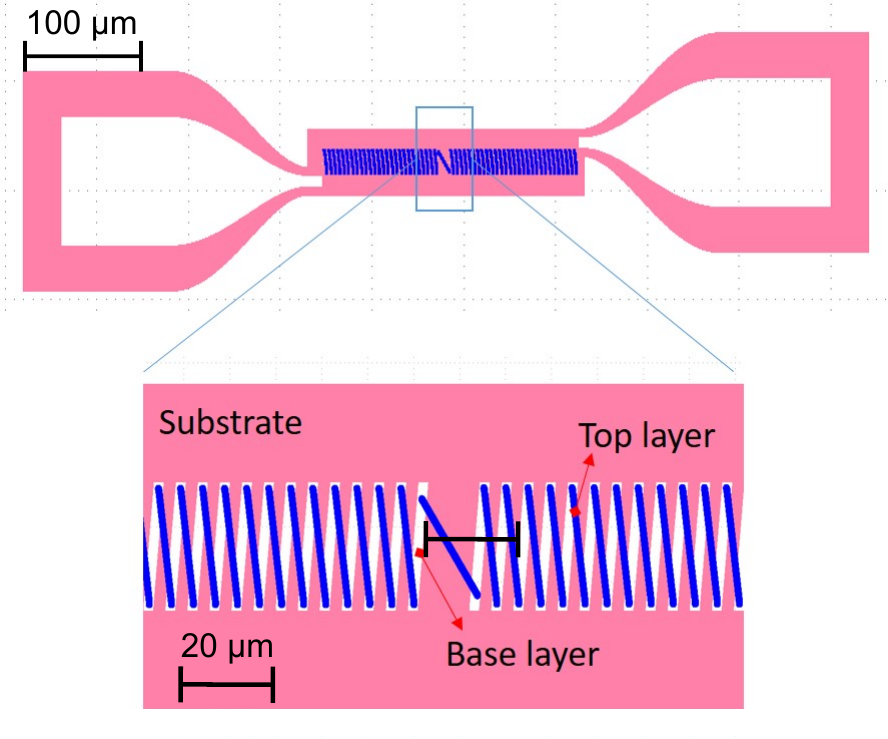}%
	\hfil
	\caption{CAD designs of flux coil representing the substrate (in pink), the metallic base layer (in white), and the metallic top layer (in blue). The design is characterized by two coils connected in series with a distance of $10\;\mu m$.}
	\label{Design_coil}
\end{figure}
Given that the ferromagnetic layer in the SIsFS junction requires an in-plane magnetic field to modify its magnetization state, the challenge is how to apply magnetic fields on-chip of the order of $5\;mT$. This value is compatible with typical hysteresis shift measured in SIsFS JJs studied so far, and to achieve a reasonable critical current levels separation to reach a qubit frequency tunability of the order of at least $20-30\%$ (Figure~\ref{fig:pattern_quad_rettang}), i.\,e., compatible with asymmetric split-transmon frequency tunability already available in literature~\cite{Hutchings2017}.

One approach to generating an in-plane magnetic field for the JJ involves using a Superconducting Coplanar Waveguide Flux line (SCPWF), directly beneath the SIsFS JJ (Figure~\ref{fig_SCPWF} (a)). A $100\;nm$ insulating layer is deposited between the flux line and the JJ to prevent galvanic contact. The CPW is designed with a width of $12\;\mu m$ and a gap of $5\;\mu m$. Due to the SCPWF's width, which is significantly larger than the JJ, as well as its proximity to it, the magnetic field within the junction area is uniform, as shown in Figure~\ref{fig_SCPWF} (b). The SCPWF is impedance-matched to $50\;\Omega$, connected to an RF signal on one side, and shorted to ground on the other. Maxwell 3D simulation using Ansys indicates that the SCPWF can sustain up to $85\;mA$ of current, which can produce an in-plane magnetic fields exceeding $4.5\;mT$ near the flux line. One possible implementation involves etching the substrate according to the SCPWF design and then depositing aluminum. This process creates a configuration where the CPW gaps are filled with the substrate (Figure~\ref{fig_SCPWF} (c)), resulting in a uniform surface across the chip suitable for qubit fabrication. Next, a $100\;nm$ insulating layer of AlO${}_x$ may be deposited on one side of the sample, specifically where the qubit will be built. This selective deposition is necessary to leave the SCPWF and its ground plane uncoated, enabling proper wire bonding.
\begin{figure*}[t]
	\centering
	\includegraphics[width=1.6\columnwidth]{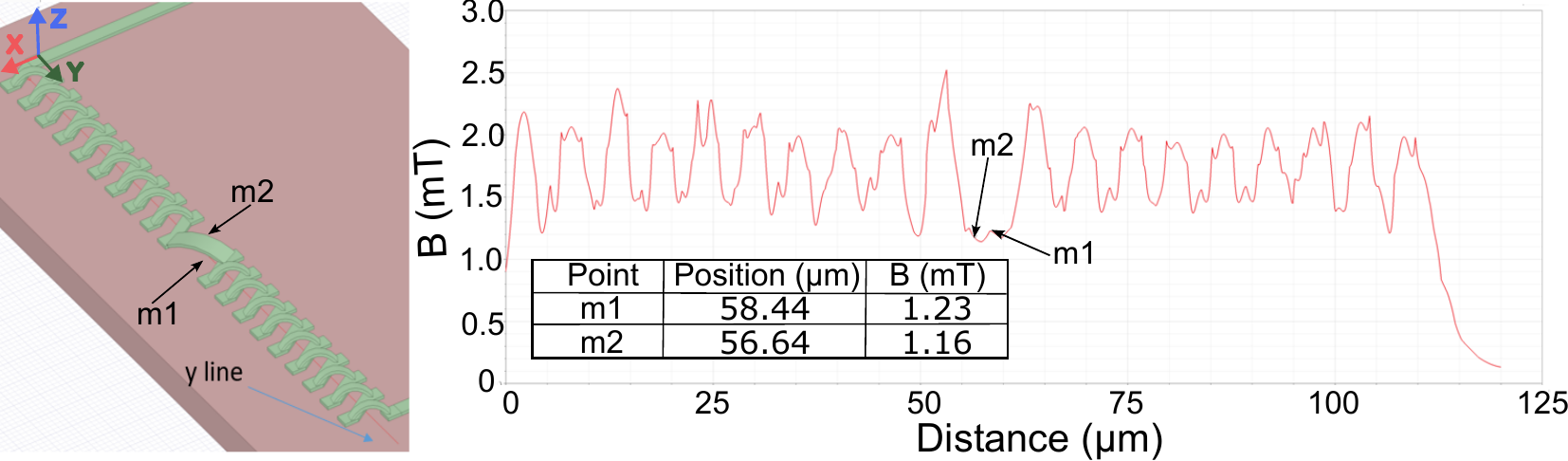}%
	\hfil
	\caption{Maxwell3D simulation of the on-chip flux coil design with the Silicon (Si) substrate (in brown) and $1 \mu m$ loop height. The graph on the right shows the magnetic field variation along they-line arrow in the left . The magnetic field is analyzed at the points m1 and m2 at the center of the coil, where the junction will be placed. The distance between m1 and m2 is around $2 \mu m$, and the magnetic field shows reasonable uniformity for a nanometric junction.
	}
	\label{Simulation_coil_2}
\end{figure*}

Another potential solution involves designing a "Helmholtz flux coil" with 3D spirals on both sides of the SIsFS junction to generate a precise and sufficiently strong in-plane magnetic flux. While directly placing a superconducting strip close the SIsFS junction could connect the qubit to a high-bandwidth environment, risking coherence loss, the Helmholtz coil design offers a safer alternative. Indeed, the presence of insulating layers aiming at electrically decoupling the qubit from the SCPWF line, increases the chance to enhance distributions of two-level systems defects, which can enter in resonance with the qubits can induce spurious relaxation and limit the coherence of the device~\cite{shnirman2002,goetz2017,Siddiqi2021}. Moreover, in order to generate magnetic fields of the order of $4.5\;mT$, the SCPWF requires larger bias currents than those typically applied through the flux lines of standard flux-tunable transmon, thus possibly contributing to detrimental localized heating effects. As a reference, applied currents through the SCPWF of the order of $25\;mA$ as in standard transmon circuits~\cite{Ahmad2024Q} may generate magnetic fields of the order of just $1.3\;mT$ in a ferrotransmon. We here demonstrate that on-chip flux coils require less bias current to generate in-plane magnetic fields.

Figure~\ref{Design_coil} presents the CAD designs of the flux coil, with the substrate layer (in pink), the metallic base layer (in white), and the metallic 3D bridges, or the top layer (in blue) highlighted. The design includes two connected coils separated by a $10\;\mu m$ distance, maintaining electrical continuity with the junction between the coils, under the bridge connecting them.
\begin{figure}[t!]
\centering
\includegraphics[width=1.0\columnwidth]{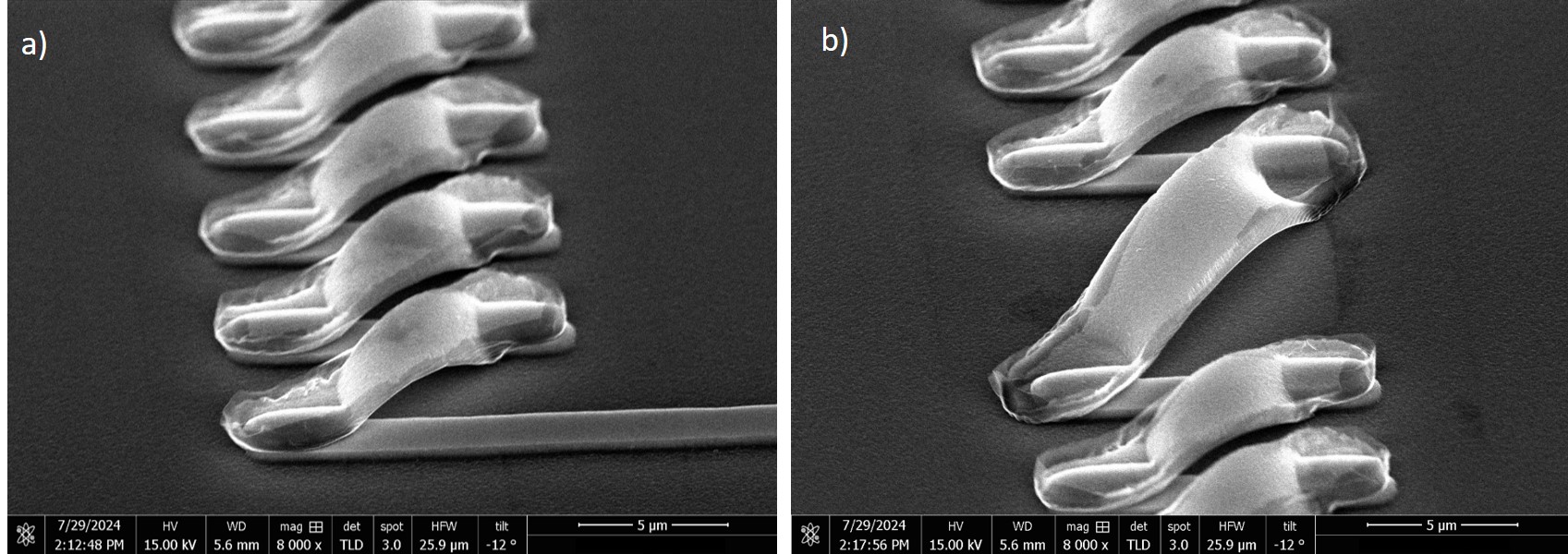}%
\hfil
\caption{a) SEM image of Helmholtz flux coil design; b) SEM image with detail on the central bridge.}
\label{SEM_coil}
\end{figure}

We simulated the flux coil using Maxwell3D HFSS simulation software to gather information about the generated magnetic field as a function of the input bias current. The goal is to magnetize the ferromagnetic material while avoiding significant dissipation on the chip.
The geometric dimensions of the flux line in Figure~\ref{Design_coil} are designed to be $2.5\;\mu m$ in width and $150\;nm$ thick, while the loop presents a width of $4\;\mu m$ and a height ranging from $1\;\mu m$ to $ 3\;\mu m$. We have performed electromagnetic simulations as a function of the loop height including the silicon substrate. While for $3\;\mu m$ loop height, we achieved magnetic fields of the order of $400\; \mu T$ in the region between the coils, reducing the loop height to $1\;\mu m$, we can get higher magnetic field values, obtaining $1.2-1.4\;mT$, corresponding to bias currents of $10\;mA$ (Figure~\ref{Simulation_coil_2}), a factor two lower than the SCPWF design.

To test the simulations, and being confident in the promising performances of flux coils on chip in terms of coherence, we fabricated a batch of flux coil samples by optimizing the base and top layers' exposure parameters and geometric dimensions. Various geometries were tested, ranging the width and gap of the flux coil from $1$ to $4\;\mu m$, and the length of the loop from $4$ to $15\;\mu m$. 

The fabrication process of the flux coil starts with a $150\; nm$ thick NbTiN layer deposited on a silicon wafer. Then a standard lithography and etching process defines the base layer. Subsequently, two lithography steps and a reflow process at elevated temperatures were employed to create the 3D bridge structures with rounded profiles. The process concluded with the deposition of a $500\; nm$ thick aluminum-titanium-aluminum stack. Figure~\ref{SEM_coil} (a) presents an SEM image of the completed flux coil, featuring a loop design with a $4\;\mu m$ width and $1\;\mu m$ height. Detailed views of the central 3D bridges are also included in Figure~\ref{SEM_coil} (b).

We tested the resistances of the devices at room temperature by using a 2-point probe station, obtaining resistance values ranging from $0.5$ to $2\;k\Omega$. This range of values correlates well with the expected sweeps of the geometric dimensions of the loops in our tests. Specifically, we observed that the resistance increases as the loop length increases, and as the wire width is reduced. The correlation between loop length and resistance is detailed in Table~\ref{Tab}, where we present room temperature measurements from four repetitions of the same geometry, labeled as "I" - "V", on the same chip.  The data underscore the importance of precise geometric control in minimizing resistance variability within these circuits. Finally, measurements at cryogenic temperatures have established the flux coils' critical current value ranging from $11.5$ to $14\;mA$, confirming the ability to provide the in-plane magnetic field from the simulation's input current. 
\begin{table*}[t]
\caption{Room temperature resistance for four repetitions of the same flux coil geometry, and where each loop length is labeled with Roman numbers "I" to "V".}
\centering
\begin{tabular}{|c||c||c||c||c||c|}
\hline
 \#1 & Loop length [$\mu m$] & Resistance [$k\Omega$] &  \#2 & Loop length [$\mu m$] & Resistance [$k\Omega$] \\
\hline
 I & $4$ & $0.50 \pm 0.01$& I & $4$ &$0.50 \pm 0.01$\\
II & $5$ & $1.03 \pm 0.01$& II & $5$ & $1.01 \pm 0.01$\\
III & $10$ & $1.63 \pm 0.01$& III & $10$ &$1.61 \pm 0.01$\\
IV & $12$ & $1.98 \pm 0.01$& IV & $12$ & $1.75 \pm 0.01$\\
V & $15$ & $2.51 \pm 0.01$& V & $15$&$2.33 \pm 0.01$\\
\hline
\#3 & Loop length [$\mu m$] & Resistance [$k\Omega$] &  \#4 & Loop length [$\mu m$] & Resistance [$k\Omega$] \\
 \hline
 I & $4$ & $0.49 \pm 0.01$ & I & $4$ &$0.52 \pm 0.01$\\
II & $5$ & $1.02 \pm 0.01$& II & $5$ & $1.12 \pm 0.01$\\
III & $10$& $1.62 \pm 0.01$& III & $10$ &$1.82 \pm 0.01$\\
IV & $12$ & $1.90 \pm 0.01$& IV & $12$ & $2.04 \pm 0.01$\\
V & $15$ & $2.46 \pm 0.01$& V & $15$&$2.5 \pm 0.01$\\
\hline
\end{tabular}
\label{Tab}
\end{table*}

\section{Conclusion}

We have explored two distinct approaches for applying an in-plane magnetic field to tune hybrid ferromagnetic transmon qubits frequency: the superconducting coplanar wave flux-line and the Helmholtz flux coil. The SCPWF, while more standard from the fabrication point of view, may introduce additional challenges due to its capacitive coupling with the qubit, which may contribute to increased decoherence. This potential for decoherence will require further investigation in future experiments. On the other hand, the Helmotz flux coil solution is expected to not significantly impact decoherence, making it a more favorable option for maintaining qubit performance. Thus, both methods have their advantages and trade-offs, each of them emerging as promising candidates for different applications: the former when coherence plays a more marginal role, and the latter when decoherence mechanisms must be reduced. Furthermore, to be compliant with reasonable qubit frequency tunability, and in order to increase the efficiency of tuning the JJ critical current, we are exploring solutions for doping through co-deposition techniques the Py ferromagnetic layer with Nb non-magnetic inclusions. The goal is to reduce both coercive and saturation magnetic fields, thus being able in the near term to implement a first proof-of-principle solution for a ferrotransmon compatible with both the on-chip in-plane magnetic flux line proposals.

\bibliographystyle{IEEEtran}
\bibliography{Bibliography.bib}

\end{document}